\documentstyle[12pt]{article}

\newcommand{\be}{\begin{equation}}
\newcommand{\ee}{\end{equation}}

\def\psnormal{\textwidth=16cm\textheight=22cm
          \oddsidemargin=0.5cm\evensidemargin=0cm
          \topmargin=0cm\parindent=1cm}
\psnormal

\begin{document}
\pagestyle{empty}

\hspace{3cm}


\vspace{1.0cm}
\begin{center}
{\bf Dilaton-Axion hair for slowly rotating Kerr black holes
\vspace{1.1cm}

S. Mignemi and N. R. Stewart
\vspace{1.3cm}

Laboratoire de Physique Th$\acute{e}$orique, CNRS/URA 769, Universit$\acute{e}$
Pierre et Marie Curie, Institut Henri Poincar$\acute{e}$, 11, rue Pierre et
Marie Curie, 75231 Paris Cedex 05, France
\vspace{0.5cm}

\vspace{1.0cm}}

\end{center}

\vspace{0.5cm}

\noindent
Campbell {\it et al.} demonstrated the existence of axion ``hair'' for Kerr
black holes due to the non-trivial Lorentz Chern-Simons term and calculated it
explicitly for the case of slow rotation. Here we consider the dilaton coupling
to the axion field strength, consistent with low energy string theory and
calculate the dilaton ``hair'' arising from this specific axion source.

\vspace{0.7cm}
\psnormal
\psnormal
%
\mbox{}

\pagestyle{plain}
\pagenumbering{arabic}
\newpage
In low energy string theory, the classical gravitational sector is modified
by the inclusion of dilaton and axion couplings. There has been much recent
interest \cite{dg,st,ca} in the new structures which are found to emerge in
black hole solutions as a direct result of the non-minimal coupling of fields
to
Einstein gravity within string gravity.

The sparseness of black hole solutions is attributed to the ``no hair''
conjecture which limits the exterior field solutions to those required
by a local gauge invariance. Hence the known static Schwarzschild holes
characterized by the mass $M$ and Kerr-Newman and Reissner-Nordstrom,
characterized additionally by angular momentum $A$ and /or charged gauge
fields, $Q$. Attempts to enlarge the space of known solutions by explicit
dependence on hair, lead to non-trivial solutions which are however, unstable
against radial perturbations \cite{bp}.

For scalar fields in particular, it has been shown by many authors \cite{jb}
that no solutions are available for minimally coupled fields even in the
presence of non-trivial potentials \cite{sw}. A slightly different result is
obtained if non-minimal couplings are considered. This problem has been
studied in the context of supergravity and Kaluza-Klein theories \cite{gg} and
it has been shown that even when non-trivial scalar ``hair'' is present, the
scalar charge is not an independent parameter but is a function of $M$, $A$ and
$Q$. The holes can therefore still be classified in terms of these three
parameters and so the scalar ``hair'' does not violate the ``no hair''
conjecture.  Similar remarks apply to string theories.

If the mass parameter of a black hole is large enough compared to the Planck
mass, then the higher order curvature invariants predicted in the string
effective action, may be neglected outside the event horizon. In particular,
for small curvature, the non-minimal coupling of the dilaton to the Gauss-
Bonnet invariant of order $\alpha '$ in the string tension, can be neglected;
moreover, the dilaton gauge field strength coupling may be retained despite
the fact that it is also $O(\alpha ')$. This is the case discussed by
Garfinkle {\it et al.} \cite{dg} and Shapere {\it et al.} \cite{st} , where
magnetic and dyonic black hole solutions are respectively obtained.

At the lowest order in the string tension, $(\alpha ')^0$, the four
dimensional theory consists of Einstein gravity, coupled to a free dilaton
field and a dilaton-axion term:
\be
S = \int\, d^{4}x\, \sqrt{-g} \left(R - 2(\nabla \phi )^{2} - \frac{1}{3}
e^{-4\phi } H^{2} + \ldots \right) .
\ee
where we assume natural units, $h=c=G=1$. It is known \cite{bg} that for the
minimally coupled case, axionic black holes correspond to static, spherically
symmetric solutions of the Schwarzschild type with mass $M$ and a purely
topological axion charge $q$,
$$
ds^{2}=-\left(1-\frac{2M}{r}\right)dt^{2} + \left(1-\frac{2M}{r}\right)^{-1}
dr^{2} + r^{2} d\Omega \;, $$
\be
\hspace{-1cm}B_{\mu \nu}=q\frac{\epsilon _{\mu \nu}}{4\pi r^{2}}\;,
\hspace{2cm}H_{\mu \nu \lambda }= 0\; ,
\ee
where $d\Omega $ is the line element on the surface of a 2-sphere.
For a pure Kalb-Ramond field $(H=dB)$, its equation of motion, $d\,^{\ast }H=0$
implies that locally, the dual of $H$ is given by $^{\ast }H=d\sigma $, where
$\sigma $ is a free massless scalar. For any Einstein-scalar field system, the
only (static and/ or stationary) black hole solutions which exist are for
{\em constant} $\sigma $ and other scalars. The non-minimal dilaton coupling
of the string-induced action (1) does not alter this uniqueness theorem so long
as the Kalb-Ramond field is itself minimally coupled, as the exterior
derivative of a 2-form field. However, one of the novel features of string
theory is the
fact that the three form field $H_{\mu \nu \lambda }$ is not minimally coupled:
gauge and gravitational anomalies \cite{gs} are removed from the theory through
the introduction of Lorentz and Yang-Mills Chern-Simons forms thus,
\be
H = dB + \omega _{L} - \omega _{Y}
\ee
In particular, the Lorentz Chern-Simons term, $\omega _{L}$, may be written
in terms of the spin connection on the manifold,
\be
\omega _{L} =Tr\, (\omega \wedge d\omega + \frac{2}{3} \omega \wedge \omega
\wedge \omega )\;,
\ee
with a similar structure for $\omega _{Y}$ in terms of the gauge field $A_{\mu}
$. In this paper we shall focus on the significant effect of $\omega _{L}$ in
producing black hole ``hair''; thus we shall ignore the gauge field sector in
the following. It has been shown \cite{cb} that for all four dimensional
spacetimes conformal to a spacetime with a maximally symmetric 2-dim. subspace,
the Lorentz Chern-Simons form is exact (i.e., $\omega _{L}=d\beta $) so that it
does not contribute to the equations of motion. Thus for a Schwarzschild
background, solution (2) holds with constant dilaton field. However for
(stationary) rotating black holes described by the Kerr metric, the
Lorentz Chern-Simons term is non-trivial and acts as a source for axion hair by
means of the Bianchi identity,
\be
dH = Tr\, R\wedge R \neq 0\;.
\ee
This mechanism was used by Campbell, Duncan, Kaloper and Olive \cite{ck} to
generate axion field strength ``hair'' in the specific case of a slowly
rotating black hole. In the limit of small rotation where the angular momentum
$A\ll M$, the Kerr metric becomes to $O(A)$:
\be
ds^{2}=-\left(1-\frac{2M}{r}\right)\,dt^{2} +
\left(1-\frac{2M}{r}\right)^{-1}\,dr^{2} -r^{2}\, d\Omega  - \frac{4MA\sin
^{2}{\theta }}{r}\,dtd\phi \;.
\ee
The dual of the Hirzebruch signature density, $^{\ast }Tr\,(R\wedge R)$, then
gives an $O(A)$ term which acts as a source for the pseudoscalar field $^{\ast
}H=da$, the equation of motion for $a$ being,
\be
\Box a = -\frac{1}{4\,!} \frac{\epsilon ^{\mu \nu \rho \sigma }}{\sqrt{-g}}
R_{\alpha \beta \mu \nu }R_{\rho \sigma }^{\ \ \beta \alpha }\;.
\ee
Campbell {\it et al.} \cite{ck} solve eq. (7) by using Green's function
techniques: for a solution to $O(A)$ it is sufficient to use the Green's
function for the Schwarzschild metric. Moreover, the back reaction of the axion
on the metric (through the energy-momentum tensor of $a$) can be ignored in
this approximation. The result \cite{ck} is regular and finite ($r\,>\,2M$):
\be
a(r\,,\theta ) \equiv f(r) \cos \theta \;; \hspace{1cm}
f(r) = \frac{5A}{48M^{3}} \left(\frac{4M^{2}}{r^{2}} + \frac{8M^{3}}{r^{3}} +
\frac{72M^{4}}{5r^{4}} \right)\; .
\ee
In this calculation the dilaton field was neglected and effectively set to a
constant. As a result of the dilaton-axion in (1), the axion hair (8) will in
turn act as a source for dilaton hair. Here we shall explicitly derive this
effect.

{}From the action (1) the equations of motion are:
$$
R_{\mu \nu } = 2\nabla _{\mu }\phi \nabla _{\nu } \phi +
 e^{-4\phi }H_{\mu \lambda \rho }H_{\nu }^{\lambda \rho } -\frac{1}{3}g_{\mu
\nu }e^{-4\phi }H^{2}$$
\be
\hspace{2cm}-\frac{1}{3}g_{\mu \nu }\nabla _{\sigma }(e^{-4\phi }H^{\lambda
\beta \alpha } R^{\sigma }_{\ \beta \alpha \lambda }) + \frac{2}{3}\nabla
_{\sigma }(e^{-4\phi }H^{\lambda \ \alpha }_{\ \mu }R^{\sigma }_{\ \nu \alpha
\lambda })\; ,
\ee
\be
\nabla _{\lambda } (e^{-4\phi }\, H^{\mu \nu \lambda }) = 0\; ,
\ee
\be
\Box \phi + \frac{1}{3} e^{-4\phi }\, H^{2} = 0\; .
\ee
If we define,
\be
\frac{1}{\sqrt{-g}} \epsilon ^{\mu \nu \lambda \rho }Y_{\rho } = e^{-4\phi }
\, H^{\mu \nu \lambda }
\ee
then eq. (10) is satisfied ($dY=0$) at least locally, for $Y=db$. The Bianchi
identity $^{\ast }(dH) = ^{\ast } Tr\,(R\wedge R) $ now becomes equivalent to
eq. (7) after the rescaling,
\be
\partial _{\mu }b= - \frac{e^{-4\phi }}{6} \partial _{\mu } a\; .
\ee
In terms of the field $a(r,\,\theta )$, the dilaton equation of motion can be
rewritten,
\be
\Box \phi + \frac{1}{N} e^{-4\phi }\,g^{\mu \nu } \partial _{\mu } a \partial
_{\nu } a = 0\; ,
\ee
where the normalization $N\equiv 3\times 6$. From eqs. (8) and (14), we see
that the dilaton hair is an $O(A^{2})$ effect. Consider the perturbative
expansion,
\be
\phi = \phi _{o} + A^{2}\, \omega (x) + \ldots
\ee
where $\phi _{o} \equiv \,$constant ($=0$) is the solution up to $O(A^{1})$.
The scalar field $\omega (x)$ represents the non-trivial dilaton hair to $O(
A^{2})$; inserting this expansion into eq. (14) and expanding the exponential
gives,
\be
\Box \omega \simeq -\frac{1}{N\,A^{2}} \partial _{\mu }a \partial ^{\mu } a\; .
\ee
We may now use the same Green's function technique to solve for $\omega (x)$ in
eq. (16). From the result of eq. (8), the source term appearing on the right
hand side of eq. (16) is,
\be
{\cal J} (r,\,\theta ) = -\frac{1}{N\,A^{2}} \left[ \frac{(r-2M)}{r} f^{\prime
} (r)^{2} \cos ^{2} \theta + \left(\frac{f(r)}{r}\right)^{2} \sin ^{2} \theta
\right]\; .
\ee
{}From the static Green's function:
\be
\Box G(x,\,y) = \frac{\delta ^{3} (x-y)}{\sqrt{-g}}\;,
\ee
we have
\be
\omega = \int d^{3}y\, \sqrt{-g(y)}\, G(x,y) {\cal J} (y) \; .
\ee
Eq. (18) is solved to $(A^{0})$ in a Schwarzschild background for a point
source at $r_{0}=b,\, \theta _{0}=0=\phi _{0}$:
\be
\frac{1}{r^{2}} \frac{\partial }{\partial r} \left[r^{2} \left(1- \frac{2M}{r}
\right)\frac{ \partial G}{\partial r}\right] + \frac{1}{r^{2} \sin \theta
}\frac{\partial }{\partial \theta }\left(\sin \theta \frac{\partial G}{\partial
\theta }\right) = \frac{\delta (r-b) \delta (\cos \theta -1)}{r^{2}}\; .
\ee
The Green's function $G(r,\,\theta )$ can be expressed in terms of the Legendre
functions $P_{l}$ and $Q_{l}$; the details of its derivation are given in ref.
\cite{ca}. We then have,
\begin{eqnarray}
\omega & = & -\int _{2M}^{\infty } db\, \int _{0}^{\pi } d\theta _{0}\, \int _{
0}^{2\pi } d\phi _{0}\, b^{2} \sin \theta _{0}\, G(r,\theta ,\phi ,b,\theta
_{0},\phi _{0}) {\cal J} (b, \theta _{0},\phi _{0}) \\
       & = & -2\pi \int _{2M}^{\infty } db\, \int _{0}^{\pi } d\theta _{0}\,
b^{2} \sin \theta _{0}\, G(r, \theta ,b,\theta _{0}) {\cal J} (b,\theta _{0})\;
;
\end{eqnarray}
where in eq. (22), we have made use of the addition theorem of spherical
harmonics; the $\phi _{0}$ integration may be performed immediately since the
source ${\cal J}(r,\,\theta )$ does not depend on the variable $\phi $.
Substituting for $G(r,\theta ,b, \theta _{0})$ \cite{ca} we obtain,
\begin{eqnarray}
\omega & = & - \sum _{l=0}^{\infty } (\frac{2l +1}{2M}) \int _{2M}^{r} db\,
\int _{0}^{\pi } d\theta _{0}\, P_{l}(b/M-1)Q_{l}(r/M-1)P_{l}(\cos \theta )
P_{l}(\cos \theta _{0}) \nonumber \\
       &   & \hspace{4cm}\times {\cal J} (b,\theta _{0}) b^{2} \sin \theta _{0}
\nonumber \\
       &   & - \sum _{l=0}^{\infty } (\frac{2l+1}{2M}) \int _{r}^{\infty } db\,
\int _{0}^{\pi } d\theta _{0}\, Q_{l}(b/M-1)P_{l}(r/M-1)P_{l}(\cos \theta )
P_{l}(\cos \theta _{0}) \nonumber \\
       &   & \hspace{4cm}\times {\cal J} (b,\theta _{0}) b^{2} \sin \theta _{0}
\end{eqnarray}
If we consider the angular integrals first, we observe that the only non-zero
contributions in $G(r,\,\theta )$ come from the $l=0$ and $l=2$ terms of the
Legendre series, corresponding to pointlike and quadrupole sources
respectively. Thus we require the following functions,
$$P_{0}(\cos \theta )=1\,;\hspace{1.5cm} P_{2}(b/M-1)=\frac{1}{2}
\left(\frac{3b^{2}}{M^{2}} - \frac{6b}{M} +2\right)$$

$$Q_{0}(b/M-1)=\frac{1}{2} \ln \left(\frac{b}{b-2M}\right)\;;$$
\be
Q_{2}(b/M-1)= \frac{1}{4} \left(\frac{3}{M^{2}} (b-M)^{2} -1\right) \ln \left(
\frac{b}{b-2M}\right) - \frac{3}{2} \left(\frac{b}{M} -1\right)
\ee
Having substituted for these we are left with eight radial integrals to
evaluate over the domain, $r\,\in \,[2M,\,\infty )$. The integrations were
carried out using a {\sc Mathematica} program. After further algebraic
manipulation, we find that the solution to eq. (16) is,
$$ \omega (r,\,\theta )= \frac{1}{169344\,N\,M^{5}} \left[\frac{14889}{2}
\left(\frac{1}{r} + \frac{M}{r^{2}}\right) + \frac{9926 M^{2}}{r^{3}} +
\frac{9989 M^{3}}{r^{4}} + \frac{21112 M^{4}}{5r^{5}}\right.$$
$$\hspace{3cm} - \frac{15176 M^{5}}{r^{6}} - \frac{29376 M^{6}}{r^{7}} - \frac{
31752 M^{7}}{r^{8}}$$

$$ \hspace{1cm} +\; 2P_{2}(\cos \theta ) \left\{ \frac{2471 M^{2}}{r^{3}} +
\frac{2513 M^{3}}{r^{4}} - \frac{2656 M^{4}}{r^{5}} - \frac{19580 M^{5}}{r^{6}}
\right. $$
\be
\hspace{3.5cm}\left. \left. - \frac{31320 M^{6}}{r^{7}} - \frac{31752
M^{7}}{r^{8}} \right\} \, \right]\hspace{ 1cm}\; ; \; (r\,>\,2M)
\ee
where $P_{2}(\cos \theta )=\frac{1}{2} (3\cos ^{2} \theta -1)$.

Thus $\omega (r,\,\theta )$ is the $O(A^{2})$ dilaton ``hair'' around a slowly
rotating Kerr black hole. It is important to stress that the axion and dilaton
``hair'' arise without the presence of a net axion/ dilaton source; the
rotation
of the hole itself is the source through the Lorentz Chern-Simons coupling.
The shape of the dilaton field $\omega $ is shown in fig. (1): $\omega (r,\,
\theta )$ is a monotonic decreasing function of $r$ and is finite at the
horizon. The quadrupole term gives rise to a small angular dependence which is
more appreciable in the vicinity of the horizon.

Unlike the axion, the dilaton ``hair'' is associated with a charge $D$ where,
\be
D = \frac{1}{4\pi }\int d^{2}\Sigma ^{\mu } \nabla _{\mu } \phi
\ee
and the integral is taken over a two-sphere at spatial infinity. Evaluating
$D$ using $\omega $ in eq. (25) gives,
\be
D= - \frac{1}{N} \left(\frac{709}{8064}\right) \frac{A^{2}}{2M^{5}}
\ee
The negative sign implies that the dilaton $\omega $ corresponds to a long
range attractive force between weakly rotating black holes. This charge is not
however a new free parameter since it is determined by the mass $M$ and angular
momentum $A$ of the black hole. The integral of the source depends on $A^{2}$
and so this charge
vanishes as $A\,\rightarrow \,0$ in the Schwarzschild limit.

We have shown that the effective string theory can give rise to non-trivial
dilaton hair through the coupling due to the Lorentz Chern-Simons term. As in
the other cases of non-minimal coupling of a scalar field with gravitation, the
dilaton charge is not an independent parameter and so the ``no-hair''
conjecture
still holds.

In this paper, we have not considered the back reaction on the gravitational
field due to the matter fields. This would be important in order to see how the
causal structure of spacetime is modified near the horizon. A rigorous
discussion of this topic would however require an exact solution of the field
equations, which seems at the moment a highly non-trivial task.

We notice however that some exact solutions have been found for a rotating
dilaton black hole in special cases of the effective string theory \cite{hh},
but where the Lorentz Chern-Simons term is neglected.

We would like to thank B. Linet for helpful discussions. The work of N.R.S was
supported by a Royal Society Fellowship. The work of S.M was supported by a
fellowship from the Minist$\grave{e}$re de la R$\acute{e}$cherch$\acute{e}$ et
Technologie.

\newpage

\newpage
\pagestyle{plain}


\begin{description}

\item[Fig.1] Dilaton hair, $\omega N$ (in polar co-ordinates) exterior to
horizon; $M=1$.

\end{description}

\end{document}